\documentclass[12pt,preprint]{aastex}
\usepackage{epsfig}

\begin{document}
\shorttitle{Comparison of clustering properties of observed
objects and dark matter halos on different mass and spatial
scales}

\shortauthors{A.V. Tikhonov, A.I. Kopylov, S. Gottlober, G. Yepes}

\title{Comparison of clustering properties of observed objects and dark
matter halos on different mass and spatial scales}
\author{Anton V.\ Tikhonov }
\affil{St.Petersburg State University, Saint-Petersburg, Russia}
\email{ti@hotbox.ru, avt@gtn.ru}
\author{Aleksandr I.\ Kopylov }
\affil{Special Astrophysical Observatory of RAS,
 Nizhnii Arkhyz, Karachai-Cherkessian Republic, 369167, Russia;}
\email{akop@sao.ru}

\author{Stefan Gottloeber }
\affil{Astrophysical Institute Potsdam, Germany}
\email{sgottloeber@aip.de}
\author{Gustavo Yepes }
\affil{Grupo de Astrofisica, Universidad Autonoma de Madrid,
Spain} \email{gustavo.yepes@uam.es }

\begin{abstract}
We investigate the  large-scale distribution of galaxy clusters
taken from several X-ray catalogs. Different statistics of
clustering like the conditional correlation function (CCF) and the
minimal spanning tree (MST) as well as void  statistics were used.
Clusters show two distinct regimes of clustering: 1) on scales of
superclusters ($\sim40h^{-1}$\,Mpc) the CCF is represented by a
power law; 2) on larger scales a gradual transition to homogeneity
($\sim100 h^{-1}$\,Mpc) is observed. We also present the
correlation analysis of the galaxy distribution taken  from DR6
SDSS main galaxy database. In case of  galaxies the  limiting
scales of the different clustering regimes are 1)10-15
$h^{-1}$\,Mpc; 2) $40-50 h^{-1}$\,Mpc. The differences in the
characteristic scales and scaling exponents of the cluster and
galaxy distribution can be naturally explained within the theory
of biased  structure formation. We compared the density contrasts
of inhomogeneities in the cluster and galaxy distributions in the
SDSS region. The value of the density contrast should be taken
into account to reconcile the observed gradual transition to
homogeneity with the apparent presence of structures on  the
corresponding scales. The estimation of the relative
cluster-galaxy bias (comparing number of clusters in different
SDSS regions with corresponding number of galaxies) gives  the
value $b = 5 \pm 2$. The distribution of real clusters is compared
to that of simulated (model) clusters (the MareNostrum Universe
simulations). We selected a cluster sample from 500 $h^{-1}$\,Mpc
simulation box with WMAP3 cosmological parameters and $\sigma_8 =
0.8$. We found a general agreement between the distribution of
observed and simulated clusters. The differences are mainly due to
the presences of the Shapley supercluster in the observed sample.
On the basis of SDSS galaxy sample we study properties of the
power law behavior showed by the CCF on small scales. We show that
this phenomenon is quite complex, with significant scatter in
scaling properties, and characterized by a non-trivial dependence
on galaxy properties and environment.
\end{abstract}

 \vspace{1ex}

\section{Introduction}
Clusters of galaxies are the most massive virialised structures in
the Universe. Within the framework of the modern "picture of the
world" with dark matter and dark energy (cosmological constant)
the structures evolve via gravitational instability starting from
small seeds which are described by the spectrum of initial
inhomogeneities. Clusters of galaxies are perfect probes of the
matter distribution on large scales. Studying their spatial
distribution one can constrain the parameters  of the $\Lambda$CDM
model ($\Omega_m$, $\sigma_8$). Baryonic gas falls into cluster
potential wells and heats up to temperatures of  the order of
$10^7$\,K so that it emits X-rays. Clusters of galaxies were
historically identified first as overdensities in the galaxy
distribution (ACO [1] and APM [2] optical galaxy cluster
catalogs). This kind of selection leads to some spurious objects
due to projection effects. When clusters are detected according to
their X-ray flux we undoubtedly deal with deep potential wells. In
this work we investigate the distribution of galaxy clusters
selected by X-ray flux from several catalogs of X-ray clusters and
compare it with simulated clusters taken from the "MareNostrum
Universe" [3].

\begin{figure}[h]
\centerline{
\includegraphics[]{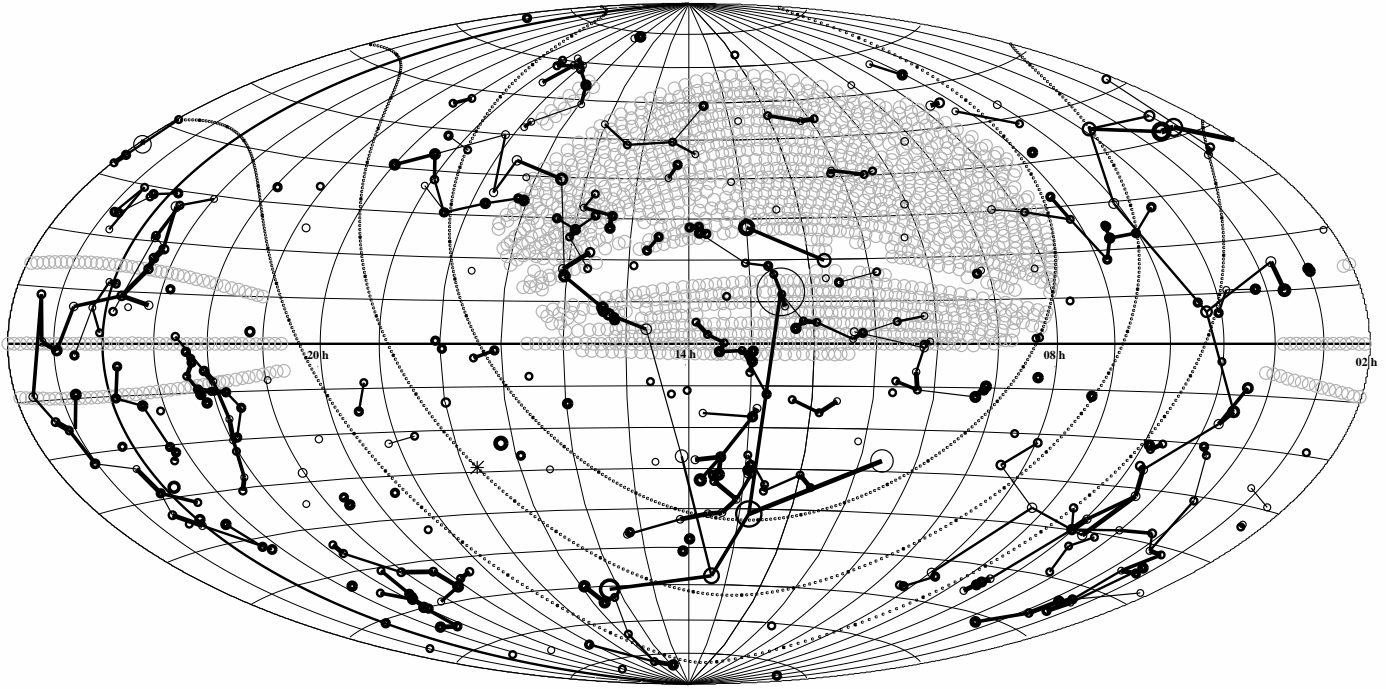}
} \figcaption{Distribution in equatorial coordinates of 400 X-ray
clusters with redshifts $z \le 0.1$ and $L_X (0.1-2.4 keV) >
1.25\cdot10^{43} h^{-2}$\,erg/s. The edges of minimal spanning
tree shorter than 45 $h^{-1}$\, Mpc are shown by solid lines.
Circles of constant galactic latitude ($b = -20^o$, $0^o$,
$+20^o$) are plotted by dotted lines. Gray circles represent the
spectral plates of the SDSS-DR6.}
\end{figure}

\begin{figure}[h]
\centerline{
\includegraphics[scale=0.8]{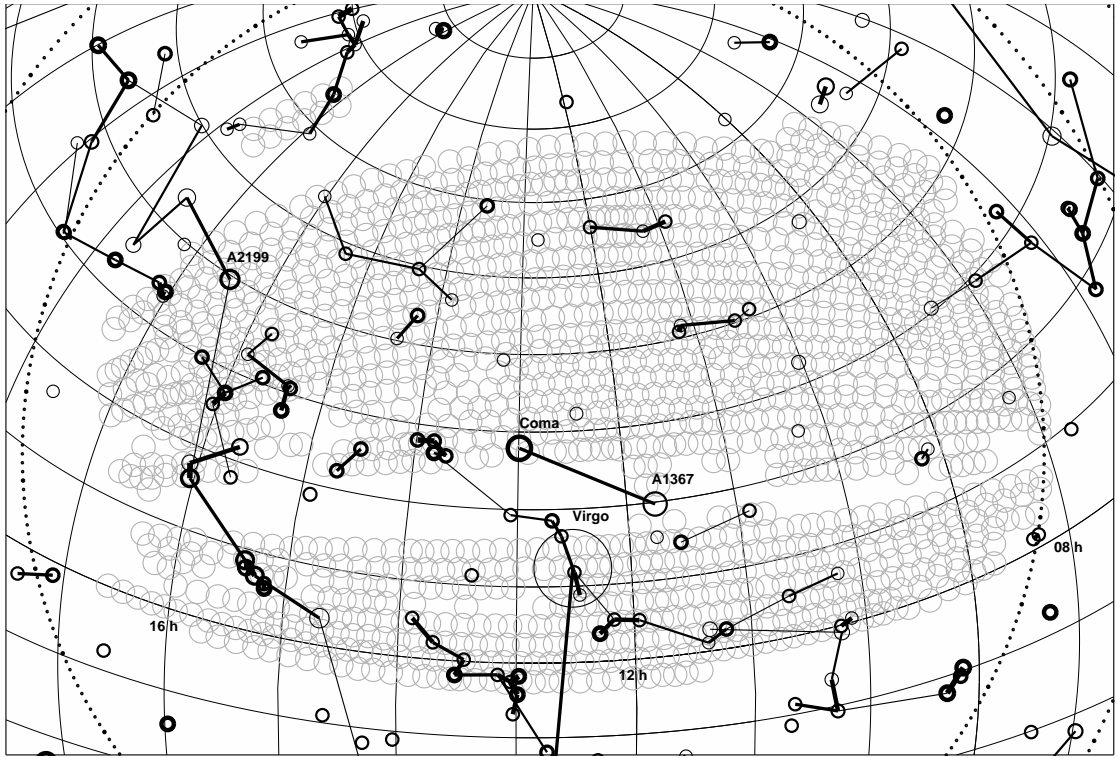}
} \figcaption{Distribution of X-ray clusters with $z \le 0.1$ and
$L_X (0.1-2.4 keV) > 1.25\cdot10^{43} h^{-2}$\,erg/s around the
Northern Galactic Pole. The edges of minimal spanning tree shorter
than $45 h^{-1}$\, Mpc are shown by solid lines. Circles of
constant galactic latitude ($b = 0^o$, $+20^o$) are plotted by
dotted lines. Gray circles represent the spectral plates of the
SDSS-DR6.}
\end{figure}

\section{X-ray galaxy cluster sample}
The X-ray cluster sample consists of the all sky ROSAT clusters
with X-ray flux $F_X \ge 3\cdot10^{-12}$\, erg/cm$^2$/s in the
(0.1-2.4 keV) energy band. Clusters were selected from the
following catalogs: 1) REFLEX (N=186 clusters) [4]; eBCS (N=108)
[5,6]; NORAS (N = 36) [7]; CIZA (N = 70, galactic latitude
$|b_{gal}| <20^o$) [8,9]. The final all-sky sample consists of 400
X-ray clusters up to redshift $z = 0.1$ with luminosities $L_X >
1.25\cdot h^{-2} 10^{43}$\, erg/s (assuming the current rate of
universal expansion - Hubble constant $H_0=100$ km/s/Mpc,
h=H/$H_0$, where H is the true value of Hubble constant). The
volume-limited sample (VL) extracted from this compilation
contains 233 X-clusters with redshifts limited by $Z_{VL} = 0.09$
(which corresponds to the radial distance of 265.3 $h^{-1}$\,Mpc
($\Omega_m = 0.24$, $\Omega_\Lambda = 0.76$) with $L_X > 2.5\cdot
h^{-2} 10^{43}$\,erg/s.

\section{Model cluster sample}
Model clusters The MareNostrum  clusters  (MN-clusters) were
extracted from the 500 $h^{-1}$\,Mpc simulation box MUWHS [10]
with cosmological parameters $\Omega_m =0.24$, $\Omega_\Lambda =
0.76$, h=0.73, $\sigma_8 = 0.8$ ($\sigma_8$, the present-day rms
mass fluctuations on spheres of radius $8h^{-1}$\,Mpc is slightly
higher than predicted by WMAP3 and in better agreement with
WMAP5). Within a box of $500h^{-1}$\,Mpc size the linear power
spectrum at redshift z = 40 has been represented by $512^3$ DM
particles of mass $m_{DM} = 8.3 \cdot 10^9h^{-1} M_{sun}$ (In the
following we assume $h=1$). The nonlinear evolution of structures
has been followed by the GADGET II code of V. Springel [11].
Clusters were identified  in the simulation by the FOF
(friend-of-friend) algorithm. For comparison with observations we
extracted the 233 most massive clusters in  a sphere of radius
265.3 Mpc (we slightly expanded the simulation box using  the
periodic boundary conditions). We have chosen this simulated
sample by keeping its number density equal to the observed cluster
density (independent of an $L_X$ - mass relation) so that the most
massive simulated clusters correspond  to the most luminous
observed clusters. We use the 3D velocities of  the simulated
clusters and place an observer to the center of the sphere
extracted from the simulation box. Then the cluster positions were
converted to redshift space. The mass of the lightest dark matter
cluster in the simulated sample is $2.46 \cdot 10^{14} M_{sun}$.

\section{Statistics of clustering}

\subsection{Conditional correlation function}
In order to compare the X-ray and simulated cluster distributions
we use different statistical methods. The conditional correlation
function (CCF) [12,13] measures the number density of objects in
spheres of radius R averaged over the spheres around all objects
of the sample that are away from sample boundaries by more then R
Mpc (integral CCF). The clusters show three distinct regimes of
clustering (Fig.3a): 1) on scales of superclusters up to a scale
of 35-40 Mpc the CCF is represented by a power law;  2) on larger
scales a gradual transition to homogeneity is observed; 3)
starting from about 100 Mpc the CCF becomes a  constant, i.e. the
number density does not anymore decline with increasing radius of
spheres. Fluctuations on scales $> 100$\,Mpc exist but evidently
they doesn't contribute to cluster number density contrasts on
such scales. Here we reach a mean number density of clusters
(which can not be obtained on smaller scales) in our VL-sample.

\begin{figure}[h]
\centerline{
\includegraphics[scale=1.0]{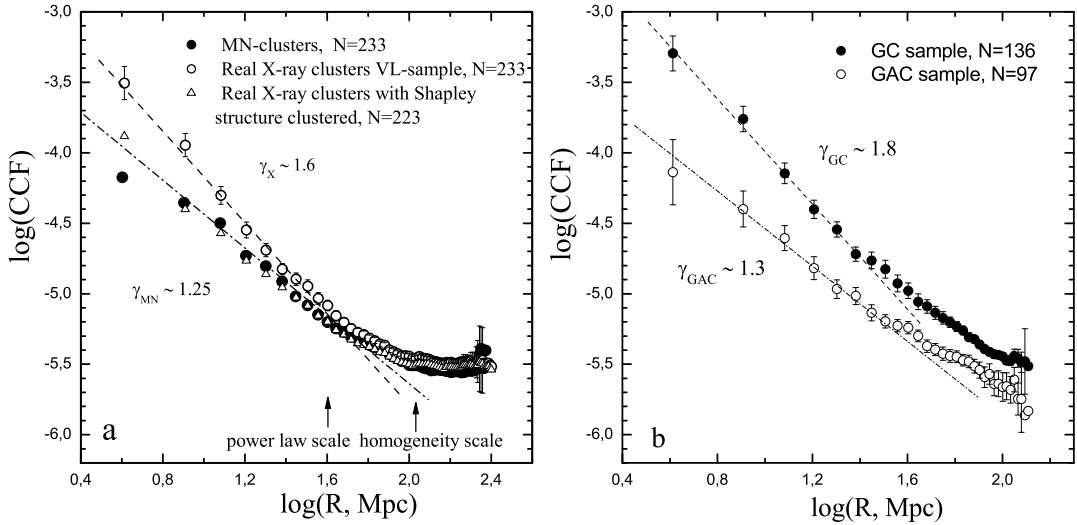}
} \figcaption{a) Comparison of CCFs for X-clusters and
MN-clusters; b) CCFs for VL cluster sample in 2 hemispheres: GAC
($123^0 < lgal < 303^0$) and GC ($303^0 < l_{gal}$  and $l_{gal} <
123^0$).}
\end{figure}

The CCFs for observed and simulated clusters look quite similar
(the second regime of clustering from $\sim40$ to $\sim100$\,Mpc
is perfectly reproduced by the simulated clusters) however the
value of the slopes on scales below 40Mpc are different: $\gamma_X
\sim1.6$ for the observed and $\gamma_{MN}\sim1.25$ for the
simulated clusters. This difference reflects the lack of close
pairs in the simulated cluster distribution with respect to the
observed ones. The comparison with a model cluster distribution
obtained from the same realization but using a smaller linking
length of the FOF algorithm (in order to identify substructure and
possible close pairs which could be linked by the original linking
length) showed that the FOF parameters have negligible influence
on the value of the slope.  In order to understand the difference
in the slopes we have gathered observed cluster pairs with
separations smaller than 5\,Mpc into single objects. This reduces
the total number of objects in the observed sample by 10.  Mostly
the clusters that belong to the Shapley supercluster (located at
$l_{gal} \sim 311^o$, $b_{gal} \sim +30^o$ and redshift $z \sim
0.05$) (Fig.6) were linked. The CCF of the reduced sample (Fig.3a)
looks nearly identical to the simulated one ($\gamma_{MN}$ is very
close to the $\gamma^p_X$ of the reduced observational sample).
When we calculate CCF separately in two hemispheres we obtain the
value of slope $\gamma_{GAC}$ very close to the model
$\gamma_{MN}$ in the GAC hemisphere that doesn't contain the
Shapley supercluster (Fig.3b). This means that the difference is
mainly produced by the Shapley supercluster -- we don't have such
a structure in the simulated cluster sample. It is an open
question how often such outstanding structures appear in the
Universe and whether they can be reproduced by $\Lambda$CDM
simulations of larger volumes which should contain larger
wavelength perturbations that could be responsible for formation
of more massive objects. Therefore, still, it is a question if we
really see in our VL-sample with CCF the universal mean X-ray
cluster number density.

\subsection{Void statistics}
We have also performed a void analysis of the same observed and
simulated samples. Starting from the largest empty spheres
non-spherical voids have been constructed by extending the
original spherical void with empty spheres of smaller radii the
center of which was inside the original void. The radius of the
smaller spheres is limited to be larger than an ad hoc parameter
0.9 of the radius of initial sphere. The process is repeated a few
times. It produces voids which are slightly non-spherical. The
mean distance between the observed (and simulated) clusters is
$~\sim 28$\, Mpc. Therefore, we have limited our voids to a
minimal radius of 20 Mpc. The cumulative void volume functions
(CVF) $\Delta V/V_{sample}$ show that the observed and simulated
voids fill the sample volumes in a similar way though in Fig.4a we
see a difference at $R_{void} \sim 80$\,Mpc: the largest simulated
voids are bigger than the observed ones. Here the differential
void function (DVF) is presented as $R_{void}$ versus rank [14]
(largest void has rank 1, $R_{void} = (3\cdot
V_{void}/4\pi)^{1/3}$). It shows rather good agreement between
observation and simulation. There are two breaks in the DVF
(Fig.4b). The one at $R_{void} \sim 45$\,Mpc (less prominent) is
associated with the scale where the power law regime of clustering
vanishes. The break at $R_{void} \sim 100$\, Mpc can be directly
associated with the scale of homogeneity. The slope of the
$R_{void} -Rank$ relation after the break is $z_v =
1/(3-\gamma_{voids})$ which gives $\gamma_{voids} \sim 1.2$ close
to the slope of the CCF on small scales.

\begin{figure}[h]
\centerline{
\includegraphics[]{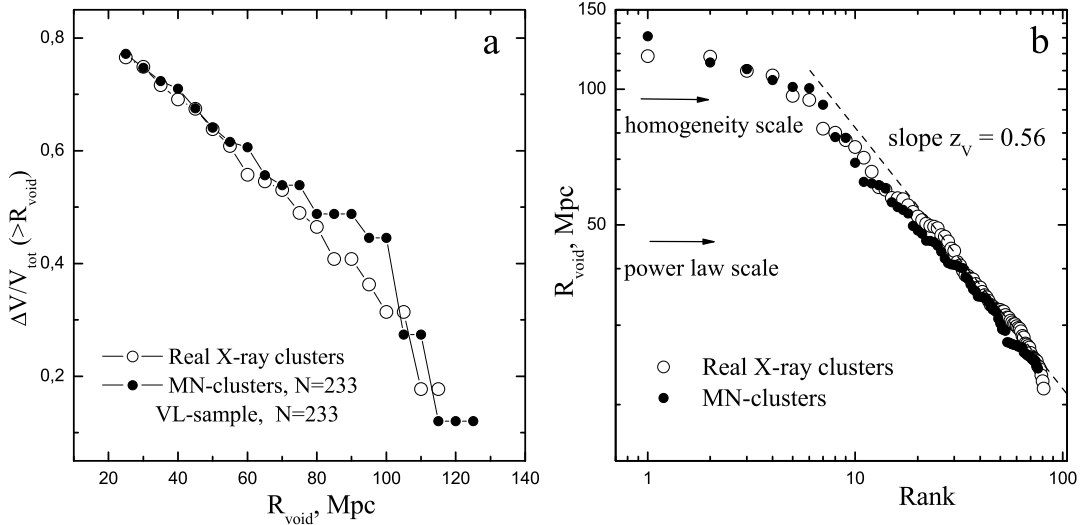}
} \figcaption{a) CVF for X-ray and  MareNostrum-clusters; b) DVF
for X-ray and  MareNostrum-clusters.}
\end{figure}

\subsection{Cross-correlation of clusters and galaxies}
From SDSS DR6 main galaxy database we selected  a region in
galactic coordinates ($48^o < l_{gal} < 210^o$, $50^o < b_{gal} <
86^o$) and built a VL sample with $z_{max} = 0.1$ and $M_{lim} =
-19.67$ (see section 5). There are 23 X-ray clusters from our
compilation in this region. In Fig.5b we present the
cluster-galaxy cross-CCF (clusters were used as centers of spheres
in which the number of galaxies inside was calculated) and compare
with the  CCF of SDSS galaxies in the selected region (star
symbols  in Fig.5b). We defined a local number galaxy contrast
$\Delta = 11/(4\pi R^3_{10}/3)/\rho_{mean} -1$, where $R_{10}$ is
the distance to the 10th neighbour ($R_{10} \sim 4$\,Mpc is the
mean value  for entire sample) and $\rho_{mean}$ is the  mean
galaxy number density in the sample. In this regard, the  clusters
located in the sample have a median galaxy constrast of  $\Delta
\sim 40$. We have randomly selected 23 galaxies separated by more
than 10 Mpc and with  $\Delta < 0$ (10 realizations) located in
void regions. The filled triangles in Fig5b show the mean cross
CCF for these low density regions. We see the same scale of
plateau on CCF at about 40 Mpc independently of the way of  the
calculation. The cluster-galaxy cross-CCF shows a stronger
correlation than observed in the entire galaxy population (CCF for
SDSS galaxies) and inherits from the cluster-cluster correlations
the length of the scaling regime. Note that all three curves in
Fig.5b, that were calculated in rather different ways converge to
the homogeneity regime at  the same scale of $\sim40$\,Mpc.

\begin{figure}[h]
\centerline{
\includegraphics[]{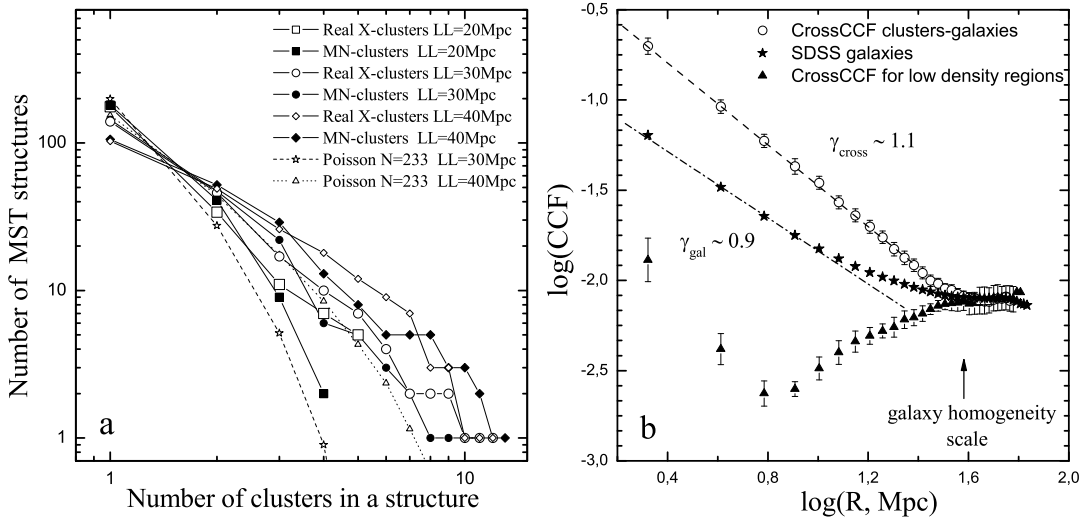}
} \figcaption{a) MST analysis for X-ray and model MN-clusters; b)
CrossCCF clusters-galaxies (open circles), CCF of SDSS galaxies
(stars; error bars are smaller than symbol size), low density
region galaxies crossCCF (triangles, 10 realizations).}
\end{figure}

\subsection{Minimal Spanning Tree}
We built the minimal spanning trees (MST) of the cluster  samples.
The MST consists of knots and edges and is constructed by
appending new knots satisfying the  condition for the distance to
the already constructed part of the tree being at a minimum [15].
The MST and void analyses give us a clue to outline the "skeleton"
of structures represented by clusters. The full length of the
truncated MST when only knots having more then 1 edge left,
normalized to the number of such knots is $L^{tr}_X = 37$\,Mpc for
the X-ray cluster sample and $L^{tr}_{MN} =38$\,Mpc for the
simulated sample (for comparison, random samples with the same
number density give after averaging $L^{tr}_R = 47$\,Mpc). Using
the MST linking lengths $LL=20$, $LL=30$ and $LL=40$\, Mpc we have
constructed the cumulative functions of structure abundances
(Fig.5a) for the observed, simulated and randomly generated
samples. Clusters in the observed sample are slightly more
structured than the simulated ones (largest differences are for
$LL=20$\, Mpc  which reflects the  same effect we   discussed
already before (section 4.1): a  lack of close pairs in the
simulated sample). But,  the largest structures in both samples
(for $LL=30$ and $40$\, Mpc)  have nearly the same number of
clusters: we see an overall agreement of abundances of observed
and simulated structures detected in  the levels of connectivity
chosen. Again we see large deviation with respect to the Poisson
sample for both values of LL which is another signature of
clustering in our samples.

\begin{figure}[h]
\centerline{
\includegraphics[scale=0.7]{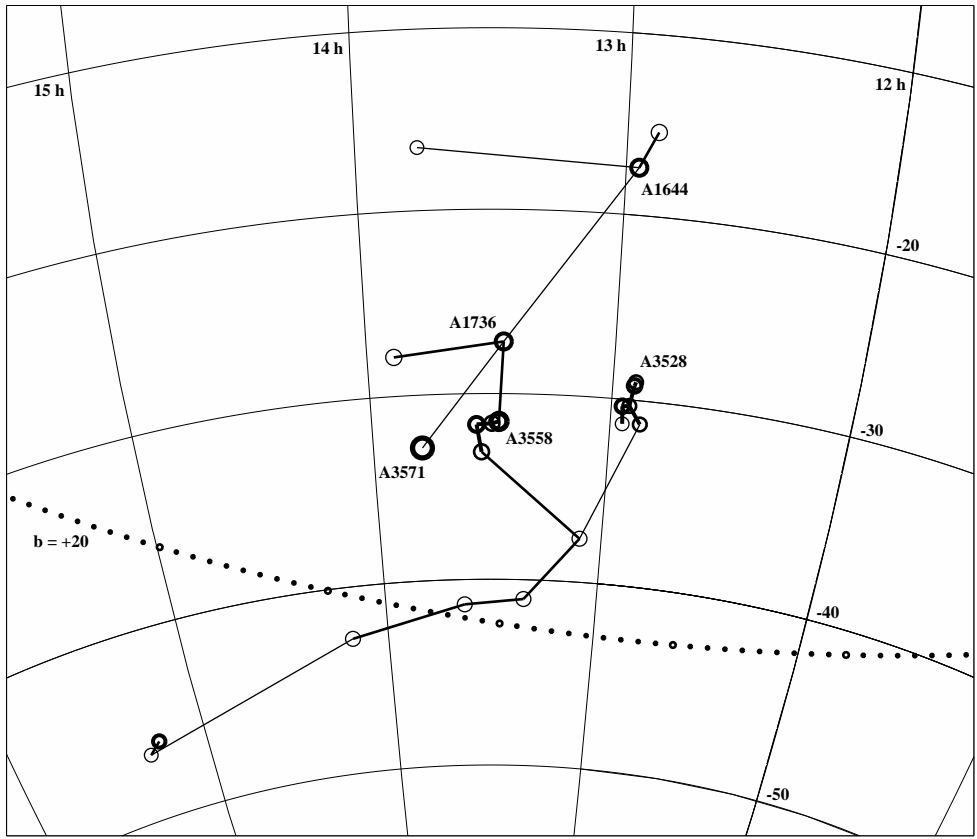}
} \figcaption{X-ray luminous ($L_X (0.1-2.4 keV) >
1.25\cdot10^{43} h^{-2}$\,erg/s) member clusters of Shapley
supercluster ($0.039 < z < 0.059$) connected by the edges of
minimal spanning tree shorter than $40 h^{-1}$\, Mpc. }
\end{figure}

\section{CCF of galaxy samples from the SDSS DR6 main galaxy
database} When analyzing the SDSS DR6 data, we have selected 3
rectangular regions from the region of spectroscopic sky coverage
for the convenience of allowance for the boundary conditions for
CCF and for ensuring sample completeness. In the ($\lambda$,
$\eta$) coordinate system of the survey, the selected regions are
S1: $48^o < \lambda < 30^o$, $6^o < \eta < 35^o$; S2: $25^o <
\lambda < 48^o$, $6^o < \eta < 35^o$; S3: $-54^o < \lambda <
-16^o$, $33^o < \eta < -17^o$.

We constructed volume limited samples to eliminate the
incompleteness in the radial coordinate. We have set the limit on
the r-band absolute magnitude Mr  for the sample of galaxies to
$M_{lim} = r_{lim}-25-5log(R_{max}(1 + z_{max})) -  K(z)$, where
$r_{lim} = 17.77$ was taken as the limiting r-band magnitude, K(z)
is the K-correction, and $R_{max}$ is the maximum value of  the
radial coordinate corresponding to $z_{max}$, ( assuming
$\Omega_\Lambda = 0.7$, $\Omega_m = 0.3$). Therefore, we have in
the VL-sample all galaxies with $M_r < M_{lim}$. The $r$
magnitudes used here were corrected for extinction. To estimate
the absolute magnitudes of the galaxies, we used an approximation
for K-correction for SDSS galaxies of the form $K(z) = 2.3537z^2 +
0.5735z + 0.18437$ [16,17]. Here, we present results for two cuts
on redshift ($z_{max}$): VL1 ($z_{max}$ = 0.12, $M_{lim} =
-20.11$) and VL2 ($z_{max} = 0.15$, $M_{lim} = -20.68$) (Fig7a,b).
\begin{figure}[h]
\centerline{
\includegraphics[]{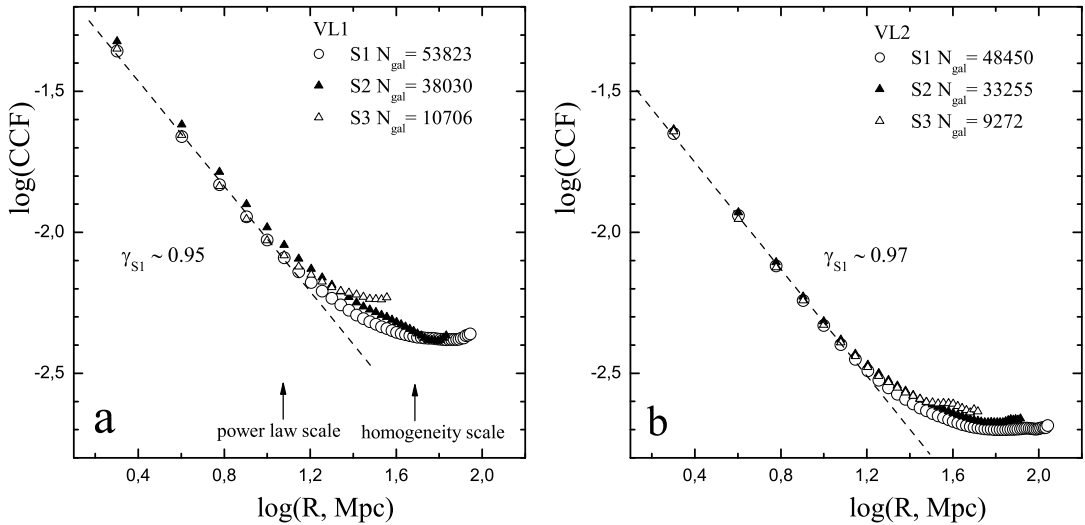}
} \figcaption{CCFs for different VL samples form DR6 SDSS main
galaxy database. Ngal - number of galaxies in a sample.}
\end{figure}

The CCF method deals with spheres that  are fully included in the
sample. For spheres with centers located at large radii they tend
to be located close to each other. We have limited our analysis to
the scale defined by the condition that spheres of large radii do
not overlap by more than half of their volumes. As can be seen in
the Figs, the power law regime is limited to scales
$\sim10-15$\,Mpc and the CCFs of different samples show rather
concerted convergence to the homogeneity regime. We should note a
small but distinct differences in the amplitudes of the
homogeneous regime in the CCFs for three different regions. This
means that,  at such scales,  we can measure the mean density with
some scatter that is caused by cosmic variance, i.e., the presence
of different structures in different samples. The characteristic
scales of galaxy correlations are significantly smaller than the
ones produced by clusters in proportions that are similar to those
obtained by early application of the more traditional two-point
correlation function $\xi(r)$ (see e.g. [18]). Such differences
have a natural explanation in the theory of biased galaxy
formation.

\section{Estimation of the relative cluster-galaxy bias on 200 Mpc
scale} The theory of structure formation predicts that the
clustering of the most massive dark matter halos (clusters) is
enhanced relative to that of the general mass distribution
(galaxies) [19, 20]. Fig.1 shows large inhomogeneities in the
distribution of clusters on scales of $100-300$\,Mpc. In the
northern region of the SDSS (Fig.2) we can estimate a relative
clustering bias (b) for the volume-limited samples of clusters
with $L_X > 2.5\cdot10^{43}$\, erg/s and galaxies with $M_r <
-19.67$. In three equal volumes of $5.2\cdot10^6$\,Mpc$^3$ defined
in the redshift intervals $0.020-0.069$, $0.069-0.087$ and
$0.087-0.100$ for the area covered by the SDSS-DR6 spectral survey
($\sim6100$ square degrees) there are 12, 32 and 13 clusters and
36612, 46785 and 38566 galaxies, respectively. These counts give a
rough estimate of cluster-galaxies relative bias $b = 5 \pm 2$ on
the scale of 200 Mpc. This estimate is consistent with the bias (b
= 3) measured for massive halos ($\sim3\cdot10^{14} M_{sun})$ in
N-body simulations, but on scales of 15-30 Mpc [21].

\section{Scaling properties of galaxies}
For samples containing galaxies with absolute magnitudes less than
M*  the galaxy integral CCF behaves  pretty much like a power law
with an exponent $\gamma_{gal} \sim 1.0$ (see section 5) up to
scales $\sim10-15$ Mpc. Galaxies more luminous than M*  tend to be
more clustered [13, 22, 23]. To investigate scaling properties of
individual galaxies on such scales, we chose all galaxies from the
sample (section 5) VL1-S1 ($z_{max}$ = 0.12) that are located more
than 10 Mpc away from the sample boundaries. For each selected
galaxy we calculated number N of neighbours in the spheres of
radii $R_{sp}$ (changing $R_{sp}$ from 2 to 10 Mpc with step 1Mpc)
and used linear approximation with slope $s_{NR}$ of
$log(N)-log(R_{sp})$ dependencies. We excluded from the analysis
very isolated galaxies and galaxies that have an error in the
slope, $\sigma_S$, larger than 0.3 (taking into account the
dispersion of slope values and excluding "bad" approximations).
After completing this procedure, about 37000 galaxies were left.
Surprisingly the mean slope $s^m_{NR} = 1.7$ (which also
corresponds to the maximum in the histogram of slope values) does
not correspond to an exponent $\gamma_{S1} \sim 1.0$ of the CCF of
the sample on scales $\sim10$\,Mpc. (For instance, in a
homogeneous scale-invariant distribution, it should be  $s^m_{NR}
= 3-\gamma_{CCF}$). The dispersion of the distribution of slopes
is significant: $\sigma_{NR} = 0.6$. By defining galaxy structures
according to their density contrast and connectivity (using the
abovementioned MST technique) we came to the conclusion  that it
is hard to associate galaxies with slopes in  a certain narrow
range with identified structures - there is a complex mixture  of
slopes with significant dispersion, but the possibility to trace
structures by their scaling properties is still an open question.
We found some amount of slopes $s_{NR} > 3$: all of them are
associated with galaxies in relatively low density regions with
$\delta < 10$ (see section 4.3). Following the approach of [24] we
performed a multi-scaling analysis that weights high and low
density regions in different way according to the positive or
negative counts-weighting exponent q-1. The generalized dimensions
$D_q$ differs significantly for different values of q ($D_q$
increases from $D_{-4} = 1.1$ to $D_2 = 1.8$) which is another
signature of the complexity of  the scaling properties on small
scales. Usually this effect is interpreted as a manifestation of
multifractality. It is evident that   galaxy clustering  on small
scales can not be described by simple models like the ones with a
unique scaling exponent.

\section{Conclusions}
The application of different complementary statistics to samples
of observed and simulated  clusters of galaxies, chosen in a way
to fit the observed number density of the volume limited X-ray
cluster sample, show general agreement in the distribution of most
luminous virialised objects in the Universe and most  massive
halos in cosmological simulations. Based on the CCF we found the
same scale ($\sim100$\,Mpc) of statistical homogeneity (in a sense
in which we  have a definite, but fluctuating, mean number density
of objects on such scales) for observed and simulated clusters.
This scale can be related to the comoving scale of the largest
wavelength of the acoustic oscillation of the photon-baryon plasma
before recombination [25]. It is very interesting to note that the
second (transitional) clustering regime (beyond the characteristic
scale of superclusters)  shown both by the observed and simulated
CCF happen to be at the same scales  of $\sim40-100$\,Mpc. The
Shapley supercluster strongly affects the value of the CCF slope
of clusters on small scales and it is responsible for the
differences in the distribution characteristics of observed and
simulated clusters: we see a lack of close massive cluster pairs
in simulations. Larger computational boxes are necessary to find
out Shapley-like structures in simulations. The MST analysis shows
that the observed clusters are slightly more structured than the
simulated ones. The observed and simulated void functions agree
except for the largest voids. In summary, the distribution of most
massive $\Lambda$CDM dark matter halos show a reasonable agreement
with the distribution of most luminous X-ray clusters of galaxies.
The significant differences in the characteristic scales of the
distribution of X-ray clusters and SDSS galaxies (power law scales
at  40 Mpc  and  10-15 Mpc and homogeneity scales at $\sim100$ and
40-50 Mpc, respectively) are similar to differences obtained in
earlier works using the two-point correlation functions. They can
be explained by the theory of biased structure formation. Our
estimation of the relative cluster-galaxy bias value ($b \sim 5$)
is in general agreement with the theoretical prediction. The power
law behaviour of the decline of the galaxy density with distance
indicated by the CCF (correlation exponent $\gamma$) on small
scales has a very complex nature: a dependence of $\gamma$ on
colors and luminosities of galaxies, a significant scatter of
individual exponents of the number-radius relation for different
galaxies in the sample, and the evidences of multifractality -
differences in the scaling properties depending on environment
(high and low density regions).

\section{Acknowledgements}
Work of A.I.K. was supported by the Russian Foundation for Basic
Research (grant 07-02-01417a). This work has been supported by the
ASTROSIM network of the European Science Foundation (ESF) (short
visit grant 2089 of A.V.T. and exchange grant 1612 of S.G.) A.V.T.
thanks the German Academic Exchange Service for supporting his
stay at  the Astrophysical Institute Potsdam. This research has
made use of the NASA/IPAC Extragalactic Database (NED). The
creation and distribution of the SDSS Archive has been funded by
the Alfred P. Sloan Foundation, the Participating Institutions,
the National Aeronautics and Space Administration, the National
Science Foundation, the US Department of Energy, the Japanese
Monbukagakusho, and the Max Planck Society. The SDSS Web site is
http://www.sdss.org/. The cluster simulations used in this
analysis  have been performed in the MareNostrum Supercomputer at
BSC-CNS (Spain) and analyzed in JUMP at Julich (Germany).

\section{References}

[1] G.O. Abell, H.C. Corwin and R.P. Olowin, A catalog of rich
clusters of galaxies, // ApJ. Suppl. Ser., Vol. 70, p. 1. (1989).

[2] G.B. Dalton, S.J. Maddox, W.J. Sutherland, G. Efstathiou, The
APM Galaxy Survey - V. Catalogues of galaxy clusters // MNRAS,
Vol. 289, p. 263 (1997).

[3] S. Gottlober, G. Yepes, Shape, spin and baryon fraction of
clusters in the MareNostrum Universe // ApJ, Vol. 664,  p. 117
(2007).

[4] H. Bohringer, P. Schuecker, L. Guzzo et al., The ROSAT-ESO
Flux Limited X-ray (REFLEX) Galaxy cluster survey. V. The cluster
catalogue // Astronomy \& Astrophysics, Vol. 425, p. 367  (2004).

[5] H. Ebeling, A.C. Edge, H. Bohringer et al., The ROSAT
Brightest Cluster Sample - I. The compilation of the sample and
the cluster log N-log S distribution
// MNRAS, Vol. 301, p. 881, (1998).

[6] H. Ebeling, A.C. Edge, S.W. Allen et al., The ROSAT Brightest
Cluster Sample - IV. The extended sample // MNRAS, Vol. 318, p.
333  (2000).

[7] H. Bohringer, W. Voges, J.P. Huchra et al., The Northern ROSAT
All-Sky (NORAS) Galaxy Cluster Survey. I. X-Ray Properties of
Clusters Detected as Extended X-Ray Sources // ApJ Supplement,
Vol. 129, p. 435  (2000).

[8] H. Ebeling, C.R. Mullis and R.B. Tully, A Systematic X-Ray
Search for Clusters of Galaxies behind the Milky Way // ApJ, Vol.
580, p. 774 (2002).

[9] D.D. Kocevski, H. Ebeling, C.R. Mullis and R.B. Tully, A
Systematic X-Ray Search for Clusters of Galaxies behind the Milky
Way. II. The Second CIZA Subsample // ApJ, Vol. 662, p. 224
(2007).

[10] G. Yepes, R. Sevilla, S. Gottlober, J. Silk, Is WMAP3
normalization compatible with the X-ray cluster abundance? // ApJ
Letters,  Vol. 666,  p. L6 (2007).

[11] V. Springel, The cosmological simulation code GADGET-2 //
MNRAS, Vol. 364, p. 1105 (2005).

[12] Tikhonov, A. V.; Makarov, D. I.; Kopylov, A. I.,
Investigation of clustering of galaxies, clusters and
superclusters by the method of correlation Gamma-function //
Bulletin of the Special Astrophysical Observatory of RAS, Vol. 50,
p. 39; ( 2000) arXiv:astro-ph/0106276.

[13] A. V. Tikhonov, Correlation properties of galaxies from the
Main Galaxy Sample of the SDSS survey // Astronomy Letters, Vol.
32, p. 721 (2006); arXiv:astro-ph/0610643.

[14] J. Gaite, Zipf's law for fractal voids and a new void-finder
// The European Physical Journal B (EPJB), 47, 93, (2005).

[15] J. Barrow, S. Bhavsar, and D. Sonoda, Minimal spanning trees,
filaments and galaxy clustering // MNRAS, 1985, Vol. 216, p.17.

[16] C. Hikage et al., Fourier Phase Analysis of SDSS Galaxies //
PASJ, Vol. 57, p. 709, (2005).

[17] M. Blanton, J. Brinkmann, I. Csabai et al., Estimating
Fixed-Frame Galaxy Magnitudes in the Sloan Digital Sky Survey //
Astron. J. 125, 2348 (2003).

[18] A.A. Klypin, A.I. Kopilov, Pis'ma v Astronomicheskiy Zhurnal
// Vol. 9, No. 2, p. 75 (1983) (in Russian).

[19] N. Kaiser, On the spatial correlations of Abell clusters //
ApJ, Vol. 284, p. L9 (1984).

[20] R.K. Sheth, G. Tormen, Large-scale bias and the peak
background split//  MNRAS, Vol. 308, p. 119 (1999)

[21] A.R. Wetzel, J.D.Cohn, M. White et al., The clustering of
massive halos
// ApJ, Vol. 656, p. 139 (2007)

[22] Zehavi I., Zheng Z., Weinberg D. et. al., The Luminosity and
Color Dependence of the Galaxy Correlation Function
// ApJ., Vol. 630, p. 1, (2005)

[23] A. V. Tikhonov, Voids in the SDSS Galaxy Survey // Astronomy
Letters, Vol. 33, No. 8, p. 499 (2007).

[24] P. Grassberger and Procaccia I., Characterization of strange
attractors // Phys. Rev Letters, 50, 346, (1983)

[25] D. J. Eisenstein, I. Zehavi, D. W. Hogg et al., Detection of
the Baryon Acoustic Peak in the Large-Scale Correlation Function
of SDSS Luminous Red Galaxies // ApJ, Vol. 633,  p. 560 (2005)

\end{document}